\documentclass[fleqn,usenatbib]{mnras}

\usepackage{newtxtext,newtxmath}

\usepackage[T1]{fontenc}

\DeclareRobustCommand{\VAN}[3]{#2}
\let\VANthebibliography\thebibliography
\def\thebibliography{\DeclareRobustCommand{\VAN}[3]{##3}\VANthebibliography}

\usepackage{graphicx}	
\usepackage{amsmath}	

\title[Serpens OB2]{On the Origin of Kinematic Structure in the Young Association\\ Serpens OB2}

\author[M. A. Kuhn et al.]{
Michael A. Kuhn,$^{1}$\thanks{E-mail: m.kuhn@herts.ac.uk (MAK)}
Robert A. Benjamin,$^{2}$
Simran S. Singh$^{1}$
\\
$^{1}$Centre for Astrophysics Research, Department of Physics, Astronomy, and Mathematics, University of Hertfordshire, Hatfield, AL10 9UW, UK\\
$^{2}$Department of Physics, University of Wisconsin--Whitewater, 800 W Main St, Whitewater, WI 53190, USA
}

\pubyear{2025}

\begin{document}
\label{firstpage}
\pagerange{\pageref{firstpage}--\pageref{lastpage}}
\maketitle

\begin{abstract}
The Serpens~OB2 association ($\ell \sim 18\fdg5$, $b\sim1\fdg9$, $d=1950\pm30$~pc) is a large star-forming complex $\sim$65~pc above the Galactic midplane, with a clumpy, elongated structure extending $\sim$50~pc parallel to the plane. We analyse probable association members, including OB stars and low-to-intermediate-mass young stellar objects (YSOs) from the SPICY catalogue. We use $^{13}$CO MWISP data to trace the molecular clouds. The OB stars are concentrated toward the centre of the association, coincident with a gap in the molecular clouds, and toward the side nearest the Galactic plane. The YSOs are distributed throughout the association, but cluster around molecular-cloud clumps. Using Gaia~DR3 proper motions to probe the association's internal kinematics, we find aligned stellar velocities on length scales $\lesssim$2~pc, two-point statistics that show increasing velocity differences and predominantly divergent motions at larger separations, and distinct velocities for star clusters within the association. Finally, the association exhibits gradual but statistically significant global expansion perpendicular to the Galactic plane, with a spatial gradient of $0.10\pm0.02$~km~s$^{-1}$~pc$^{-1}$. The clumpy stellar distribution, correlated velocities on small scales, and increasingly divergent motions on larger scales are consistent with an initial velocity field inherited from a turbulent molecular cloud modified by stellar feedback. The global vertical expansion may arise from large-scale turbulence or feedback-driven shell expansion, with the H\,{\sc ii} region Sh~2-54 preferentially pushing the molecular gas away from the Galactic plane. Ser~OB2 demonstrates that the multi-scale expansion of an OB association can begin even while star formation is still ongoing throughout the complex.
\end{abstract}


\section{Introduction}

Large OB associations---often containing multiple young clusters---exhibit complex kinematic substructures \citep[][]{2020NewAR..9001549W,2023ASPC..534..129W}. 
Their kinematics may reflect turbulent or collapsing natal clouds \citep{2007ARA&A..45..565M,2019MNRAS.490.3061V}, gas removal by OB-star feedback (winds, expanding H\,{\sc ii} regions, supernovae) that can drive cluster expansion or dispersal  \citep{1983ApJ...267L..97M,2000ApJ...542..964A,2003ARA&A..41...57L,2007MNRAS.380.1589B}, tidal disruption \citep{2012MNRAS.419..841K,2012MNRAS.426.3008K}, sequential star formation in expanding shells \citep{1977ApJ...214..725E,2007MNRAS.375.1291D}, and acceleration of clouds by the rocket effect \citep{1955ApJ...121....6O,2019MNRAS.487.2977G,2023A&A...679L..10P}.
Gaia studies of young star clusters in massive star-forming regions show that, once the natal gas has dispersed sufficiently for the stars to become optically visible, most individual clusters are expanding \citep{2019ApJ...870...32K,2020ApJ...899..121L,2021ApJ...917...21S,2022A&A...668A..19M,2024MNRAS.533..705W,2024A&A...692A.166A}. 

The local expansion of clusters or subclusters, however, need not imply coherent expansion of the parent OB association \citep[e.g.,][]{2018MNRAS.476..381W}.
In large complexes containing multiple expanding groups, the groups' motions relative to one another can appear largely uncorrelated \citep{2019ApJ...870...32K,2020ApJ...899..128K}. Analysing 109 OB associations, \citet{2020MNRAS.495..663W} found highly substructured velocity fields with only localised expansion, and concluded that the observed kinematics are inconsistent with simple expansion from one or more centres. In contrast, careful examination of stellar clustering in Sco--Cen, a nearby 3--19~Myr old OB association, suggests that its present-day morphology arises from multiple non-coeval subpopulations that have expanded from their birth sites \citep{2023A&A...678A..71R}, which has given rise to an overall divergent velocity field \citep{2025arXiv250913607H}. To investigate the origin of these multi-scale kinematic patterns, we examine the younger, less-studied OB association Serpens~OB2.

Ser~OB2 is a prominent OB association with active star formation located $\sim$2$^\circ$ ($\approx$65~pc) above the Galactic midplane \citep{Reipurth2008_NGC6604}. This association contains the young cluster NGC~6604 and the giant H\,{\sc ii} region Sh~2-54, which launches a $\sim$200~pc long `thermal chimney' above the Galactic plane \citep{1984KlBer..27..295R,1987A&A...183..327M}. The distance to Ser~OB2 is $1950\pm30$~pc, placing this association at the far end of the Sagittarius Spur---a kpc-long chain of star-forming regions that also includes the massive star-forming regions NGC~6530, M16, and M17 \citep{SgrSpur}. A significant fraction of nearby star-formation activity appears to arise from giant filamentary structures like this \citep{Zucker_PPVII}, so it is important to understand how such configurations of interstellar gas affect the kinematics and clustering of the stars they form. 

Owing to its distance, previous analysis of the stellar content has mostly focused on the association’s early-type members \citep[][and references therein]{Reipurth2008_NGC6604}. However, with infrared surveys, it has become possible to identify low- and intermediate-mass young stellar objects (YSOs), extending the sample of members across most of the stellar mass range. We have combined these catalogues with Gaia astrometry \citep{GaiaCollaboration2016} to investigate stellar kinematics in this region. The paper is organised as follows. Section~\ref{sec:data} describes the survey data. Section~\ref{sec:mem} presents the OB and YSO members. Sections~\ref{sec:clust} and \ref{sec:kin} examine the stellar clustering and kinematics. Section~\ref{sec:cloud} investigates the molecular cloud. Section~\ref{sec:discussion} interprets the kinematic results in terms of stellar feedback and turbulence in the natal cloud. And, Section~\ref{sec:conclusion} provides our conclusions.

\begin{figure*}
    \includegraphics[width=0.75\textwidth]{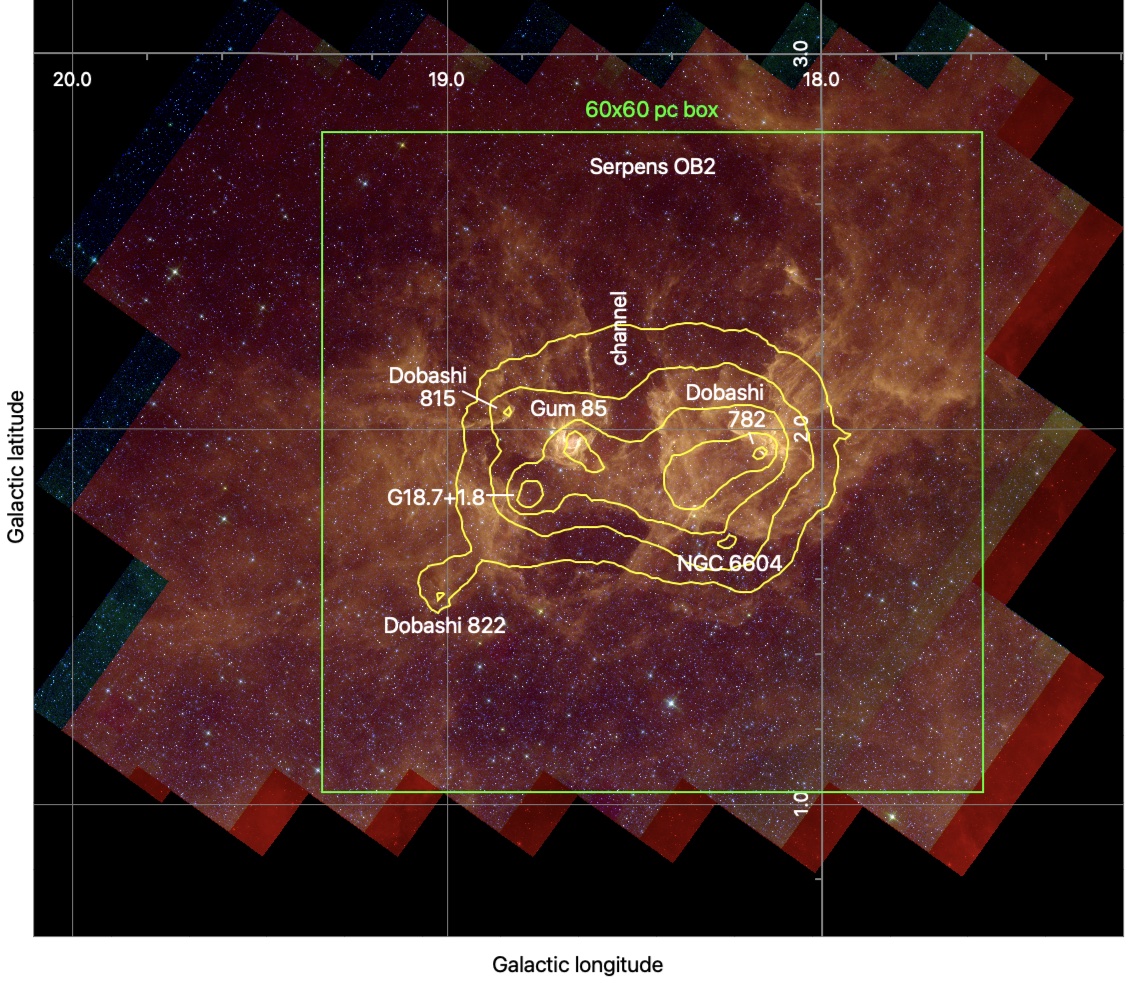}
\caption{ 
    Spitzer/IRAC mosaic of the Ser~OB2 region in the 3.6~$\mu$m (blue), 5.8~$\mu$m (green), and 8.0~$\mu$m (red) bands. The yellow contours indicate the surface-density distribution of cluster members (Section~\ref{sec:clust}), with the $60\times60$~pc box used for this analysis indicated (green rectangle). Several subregions are labelled. }
    \label{fig:irac}
\end{figure*}

\section{Data}\label{sec:data}

Ser~OB2 was observed by the Spitzer Space Telescope \citep{2004ApJS..154....1W} during the Galactic Legacy Infrared Midplane Survey Extraordinaire \citep[GLIMPSE;][]{Benjamin2003,2009PASP..121..213C}, as part of the GLIMPSE~3D programme (PI R.\ Benjamin; Program ID 30570).
The GLIMPSE~3D field containing Serpens~OB2 (Fig.~\ref{fig:irac}) sits above the Galactic midplane with dimensions $\sim$2$.\!\!^\circ75\times2.\!\!^\circ25$ centred at $(\ell,b) \approx (18.\!\!^\circ6 ,2.\!\!^\circ0)$. This strip was observed between 13--14 May 2007 in the Infrared Array Camera \citep[IRAC;][]{2004ApJS..154...10F} 3.6, 4.6, 5.8, and 8.0~$\mu$m channels, with 2~s integration times at each position. 
The GLIMPSE data reduction and catalogue construction, using point-spread-function photometry, are described by \citet{Benjamin2003} and online NASA/IPAC Infrared Science Archive (IRSA) documents.\footnote{\url{https://irsa.ipac.caltech.edu/data/SPITZER/GLIMPSE/overview.html}} 

The analysis is also based on the third Gaia Data Release \citep[Gaia DR3;][]{GaiaCollaboration2016,GaiaCollaboration2021,GaiaCollaboration2023}, including astrometry \citep{Lindegren2021_as} and photometry \citep{Riello2021}. For Ser OB2 sources with $G\sim17$~mag, Gaia's typical precisions are $\sim$0.08~mas for parallaxes and $\sim$0.09~mas~yr$^{-1}$ for proper motions. However, spatially correlated systematic errors of up to 0.026~mas and 0.023~mas~yr$^{-1}$ in parallax and proper motion, respectively, limit the precision of absolute astrometry \citep[][their Section~5.6]{Lindegren2021_as}. We apply the \citet{Lindegren2021_zp} astrometric zero-point corrections and omit analysis of sources with renormalised unit weight error $\mathrm{RUWE} > 1.4$. Furthermore, we only examine astrometry for sources with $\mathrm{astrometric\_sigma5d\_max} \leq 1$~mas to ensure that the measurements are sufficiently precise to be informative. 

\section{Membership}\label{sec:mem}

Sources whose photometry showed mid-infrared excesses consistent with circumstellar discs or envelopes were included in the Spitzer/IRAC Candidate YSO (SPICY) catalogue \citep{SPICY}, following a multi-step classification strategy. Starting with the GLIMPSE high-quality `Catalog,' cuts were applied based on IRAC colours and uncertainties to remove sources whose photometry is incapable of providing a reliable detection of infrared excess \citep[][]{Povich2013}. Next, sources that could be explained by reddened stellar atmosphere models  \citep[e.g.,][]{CastelliKurucz2003,Indebetouw2005} were rejected using spectral energy distribution fitting. Finally, a random-forest-based classifier was used to separate YSOs from intrinsically red objects, e.g., active galactic nuclei, star-forming galaxies, (post)-asymptotic giant branch stars, dusty red giant stars, cataclysmic variables, classical Be stars, and symbiotic stars. This classifier was trained on probable cluster members from massive star-forming regions \citep[][]{Feigelson2013,Povich2013,Broos2013}, while contaminants identified in these studies and sources from multiple Galactic sight lines with little-to-no star-formation activity comprised the `non-YSO' training set. The full SPICY catalogue contains $\sim$120,000 objects, of which $\sim$860 lie in the GLIMPSE 3D Ser~OB2 field. While quantitative contamination rates are challenging to estimate, preliminary follow-up studies suggest that it is low \citep[e.g.,][]{Kuhn2023}.\footnote{\citet{Kuhn2023} estimated a contamination rate $<$10\% for spatially isolated, optically visible SPICY members. Nevertheless, it has yet to be demonstrated that this contamination rate applies to the full sample.}

Candidate OB members ($M>8$~$M_\odot$) were compiled from the Alma Luminous Star (ALS) III catalogue \citep{2025MNRAS.543...63P}, an update to the ALS catalogue where sources have been vetted using distance-corrected Gaia colour-magnitude diagrams. We include ALS~III stars consistent with being massive stars, including those in their massive catalogue (labelled `M') along with ALS candidates with bad Gaia astrometry (labelled `A') or colours (labelled `C'). Out of $\sim$90 potential massive members, three quarters are in the `M' category, but several important sources, including MY~Ser, the O supergiant multiple-star system at the centre of NGC~6604, have flagged Gaia issues. One source, ALS~17499 (=~SPICY~79715), appears in both catalogues. 

We examined 11 Infrared Astronomical Satellite (IRAS) candidate YSOs from \citet{2000AJ....120.2594F}, but only IRAS~18166-1128 has a Gaia parallax consistent with Ser~OB2 membership, so we have not added these sources to our sample. In addition, \citet{2012A&A...545A..89V} proposed $\sim$60 candidate YSOs in a small region on the eastern side of Ser~OB2, based on similar datasets to those used by SPICY, but with different selection criteria. To preserve spatially uniform selection across the full Ser~OB2 field, we decided not to include additional sources from this list. 

While our list of potential members is the largest assembled for Ser~OB2 (Table~\ref{tab:sample}), our selection criteria are insensitive to disc-free pre-main-sequence stars. Nevertheless, the SPICY and ALS~III selections apply spatially uniform criteria across the GLIMPSE~3D field and provide a homogeneous set of probable members with which to compare the spatial and kinematic properties of the high- and low-mass populations.

\begin{table*}
    \centering
    \caption{Probable Association Members (OB Stars and YSOs)}
    \label{tab:sample}
    \begin{tabular}{lccrc}
        \hline
\multicolumn{1}{c}{Star Name} & \multicolumn{1}{c}{R.A.}  & \multicolumn{1}{c}{Decl.}  & \multicolumn{1}{c}{Gaia DR3} & \multicolumn{1}{c}{Group} \\
\multicolumn{1}{c}{} & \multicolumn{1}{c}{(deg)}  & \multicolumn{1}{c}{(deg)}  & \multicolumn{1}{c}{} & \multicolumn{1}{c}{} \\
        \hline
SPICY 79382 & 274.345989 & $-12.194012$ & & Ser~OB2 ridge W\\
SPICY 79385 & 274.346755 & $-11.986258$ && Ser~OB2 ridge W \\
SPICY 79387 & 274.347122 & $-12.153139$ &  4153579966233449600& Ser~OB2 ridge W\\
BD\,$-$11$^\circ$ 4581 & 274.347245& $-11.749321$ &  4153880717006159872 & Ser~OB2 ridge W\\
SPICY 79391 & 274.347603 & $-12.361940$ &  4153564057674463616& Ser~OB2 ridge W\\
SPICY 79392 & 274.347701 & $-11.985276$ &  4153679678221028480& Ser~OB2 ridge W\\
SPICY 79393 & 274.348619 & $-12.031330$ & & Ser~OB2 ridge W\\
SPICY 79394 & 274.349002 & $-12.086487$ &  4153674966596672128& Ser~OB2 ridge W\\
SPICY 79395 & 274.349055 & $-12.324524$ & & Ser~OB2 ridge W\\
SPICY 79397 & 274.349322 & $-12.169315$ &  4153579858837645056& Ser~OB2 ridge W\\
        \hline
    \end{tabular}
\flushleft{Note: R.A.\ and Decl.\ are equinox J2000.  Gaia matches are only indicated for sources meeting our astrometric quality criteria. Stars with parallaxes inconsistent with membership are not included. 
     This table is available in its entirety (899 rows) in the supplementary online material. A portion is shown here for guidance regarding its form and content.}
\end{table*}

\begin{figure*}
\includegraphics[width=0.48\textwidth]{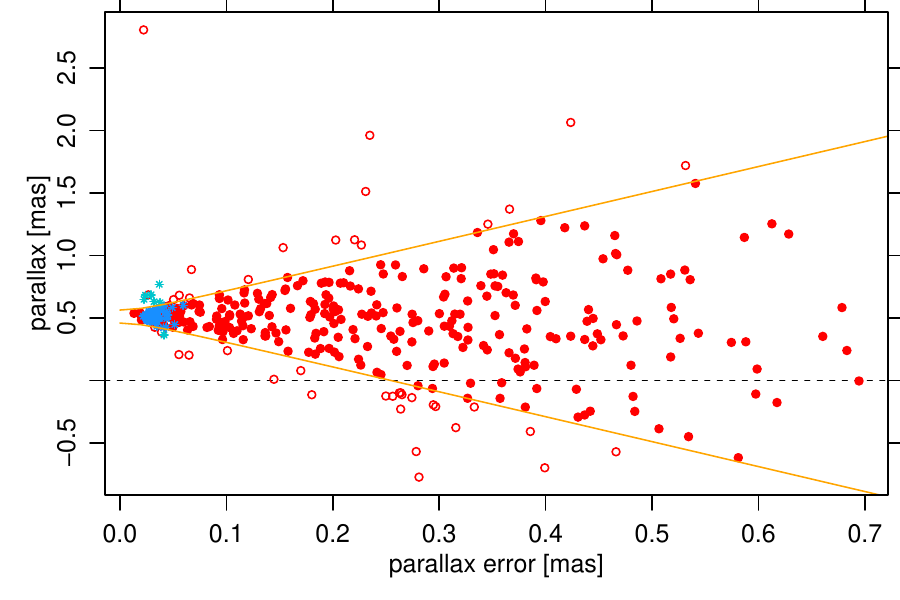}
    \includegraphics[width=0.48\textwidth]{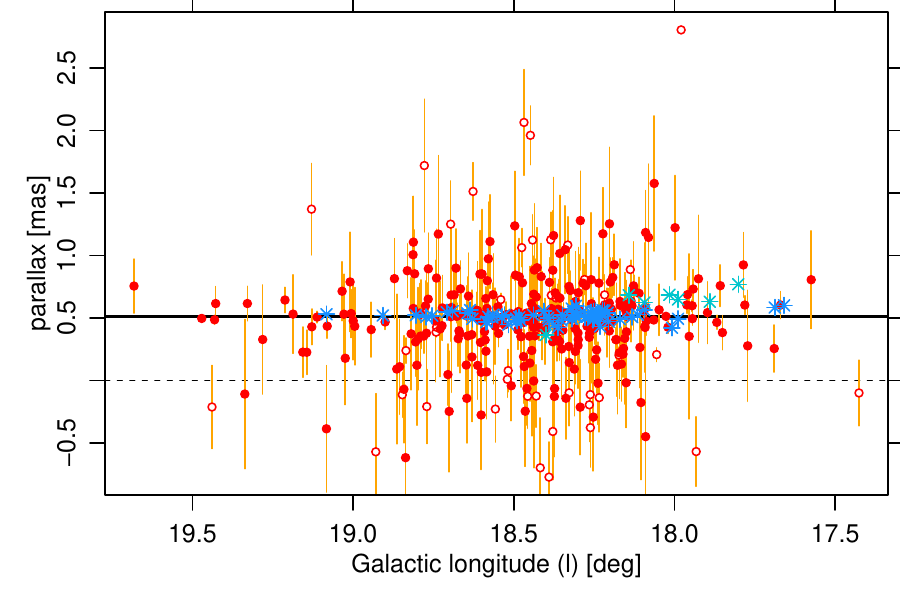}
    \caption{Parallaxes for OB stars (blue asterisks) and YSOs (red circles) in the Ser~OB2 field. Left: Corrected Gaia parallax ($\varpi$) versus formal parallax uncertainty ($\sigma_\varpi$). The diagonal lines show the $\pm2\sigma$ envelope around the association mean $\varpi_0=0.512$~mas, using the quadrature sum of statistical and systematic uncertainties. Sources whose parallaxes differ from $\varpi_0$ by more than $2\sigma$ are indicated by open circles (YSOs) or fainter symbols (OB stars). Right: Parallax versus Galactic longitude ($\ell$) using the same symbols as the left panel. The $1\sigma$ error bars are indicated for YSOs (OB-star parallax uncertainties are smaller than their symbols). The horizontal black line marks $\varpi_0$. The larger parallax spread of the YSOs reflects their larger formal uncertainties.}
    \label{fig:plx}
\end{figure*}

\subsection{Parallax Membership Refinement}

\citet{SgrSpur} used a Bayesian model to infer a mean Gaia parallax of $0.512\pm0.007$~mas ($1950\pm30$~pc) for Ser~OB2. This is consistent with an earlier very-long-baseline interferometry measurement of $0.500\pm0.019$~mas for the maser G018.34+01.76, associated with IRAS~18151-1208 \citep{2019ApJ...885..131R}. Here, we retain YSOs and OB stars as probable members only if the measurements are consistent with 0.512~mas within uncertainties. The brighter OB stars have smaller astrometric uncertainties and lie close to this value, whereas the fainter YSOs show larger scatter but remain concentrated around this mean (Fig.~\ref{fig:plx}).

Using two standard deviations as a cut-off, and accounting for both the formal parallax uncertainties and the systematic error, we reject the membership of 13\% of the YSOs and 10\% of the OB stars in the field. The remaining sources that have good Gaia astrometry include 60 OB stars and 284 YSOs. The 21 OB stars and 534 YSOs lacking good Gaia astrometry are retained as possible members; however, they cannot be used for kinematic analysis. 

Six OB stars were rejected because their parallaxes are too large for association membership. All lie on the low-$\ell$ side of the field ($17.8^\circ \lesssim \ell \lesssim 18.2^\circ$) and are $\sim$350--650~pc closer than the mean association distance. Given that Ser~OB2 spans only $\sim$50~pc in the plane of the sky, their inclusion would require an extreme elongation of the association towards the Sun. It is more likely that these OB stars are associated with foreground star-forming regions such as M17 or the Eagle Nebula, which are projected on the same side of the field. One OB star was rejected because its parallax places it $\sim$800~pc behind Ser~OB2. After excluding these seven OB stars, we find no significant $\ell$--parallax gradient across Ser~OB2 based on a Kendall’s $\tau$ correlation test ($p>0.05$).

\begin{figure}
    \includegraphics[width=0.48\textwidth]{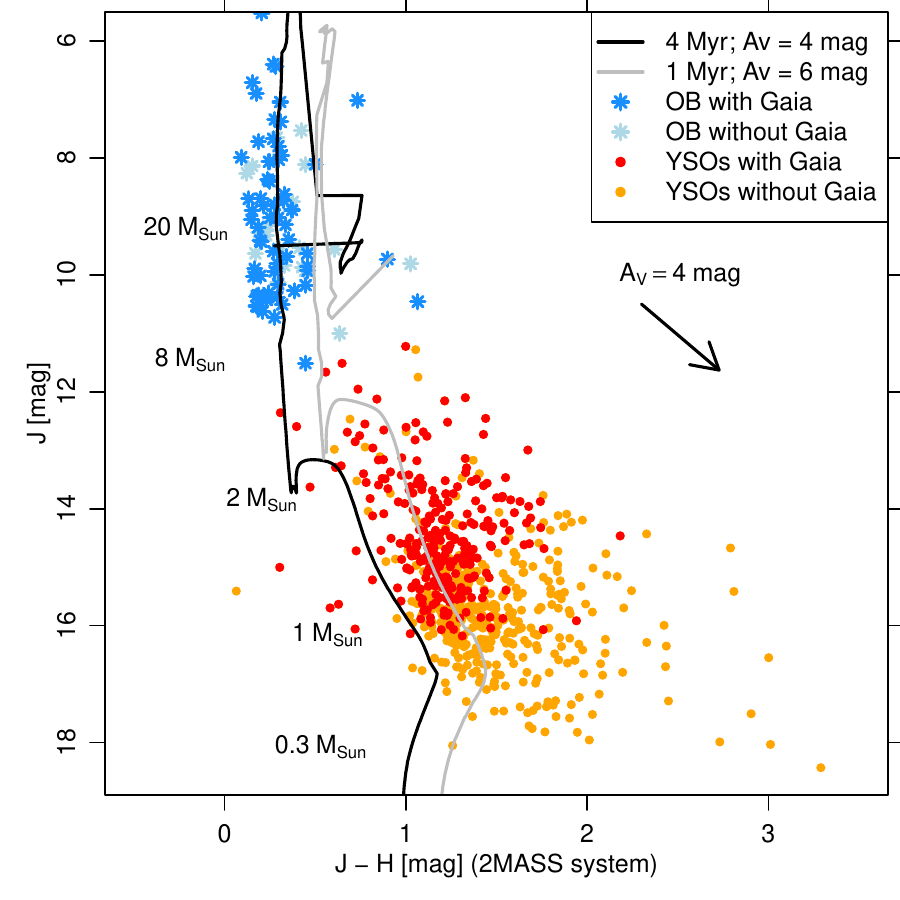}
    \caption{Near-infrared colour--magnitude diagram of probable Ser~OB2 members (OB stars and YSOs). Stars are colour-coded by the availability of high-quality Gaia astrometry. The YSO sample reaches $\sim$0.4~$M_\odot$, but the subset with Gaia astrometry is sensitive only above $\sim$1~$M_\odot$. Stars with discrepant parallaxes are excluded from the diagram. PARSEC isochrones \citep{2012MNRAS.427..127B} are shown at the Ser~OB2 distance, reddened to extinctions typical of members.}
    \label{fig:cmd}
\end{figure}

\subsection{Member Properties}\label{sec:isochrone}

Previous studies of Ser~OB2's OB content estimated stellar ages of 4--5~Myr \citep{1978AJ.....83..266F,2000A&AS..144..451B,2005A&A...438.1163K,Reipurth2008_NGC6604}. These ages are consistent with the presence of the Wolf-Rayet system, WR 113 \citep[WC8d+O8--9 IV;][]{2001NewAR..45..135V}, given that the evolution of a WC8 star requires $\gtrsim$3~Myr \citep{2005A&A...429..581M,2007ARA&A..45..177C}. 

The $J$ versus $J-H$ colour--magnitude diagram is shown in Fig.~\ref{fig:cmd}. Using PARSEC stellar evolution models \citep{2012MNRAS.427..127B} with an age of 4.4~Myr \citep{2005A&A...438.1163K}, an extinction of $A_V\sim4$~mag reproduces the OB locus but lies significantly bluer than most YSOs. For the 4.4~Myr isochrone to pass through the middle of the YSO distribution on the diagram, it would need to be reddened by $A_V \sim 10$~mag. Such an extinction is implausibly high given that many of these stars have Gaia astrometry. For example, a 2~M$_\odot$ star with this age and extinction would have a Gaia magnitude of $G=20.6$~mag, which is too faint for reliable astrometry, yet dereddening on the near-infrared diagram suggests that some YSOs are in this mass range. A younger isochrone alleviates this tension. With a 1~Myr isochrone, an extinction of $A_V \sim 6$~mag is sufficient to explain the YSO colours, and a 2~M$_\odot$ star would then have $G=18.4$~mag, within the range where Gaia provides astrometry. Owing to differential extinction in massive star-forming regions and degeneracies between effective temperature and extinction in spectral-energy-distribution fitting, it is difficult to obtain precise masses and ages for individual stars. Nevertheless, the constraints provided by the colour-magnitude diagram suggest that the YSOs are consistent with an age of $\sim$1~Myr and they are likely to be, on average, younger and more extincted than the OB population. 

\subsection{Extrapolated Total Stellar Membership}

Assuming our OB member list is complete and finding no evidence of past supernovae, we can extrapolate the total stellar population of the association by scaling from the initial mass function (IMF). The \citet{Maschberger2013} IMF predicts that OB stars ($M>8$~$M_\odot$) comprise $\sim$0.7\% of all newly formed stellar-mass objects ($M>0.08$~$M_\odot$). Given $\sim$80 probable OB stars in Ser OB2 after removing those with inconsistent parallaxes, the association is likely to have a total population of $\sim\!\!10^4$ stellar-mass members. For comparison, \citet{2022AJ....163...24L} estimates 8300 stellar-mass members in Sco--Cen, meaning that these two associations have similar stellar population numbers. 

\section{Clustering}\label{sec:clust}

\begin{figure*}
    \includegraphics[width=0.48\textwidth]{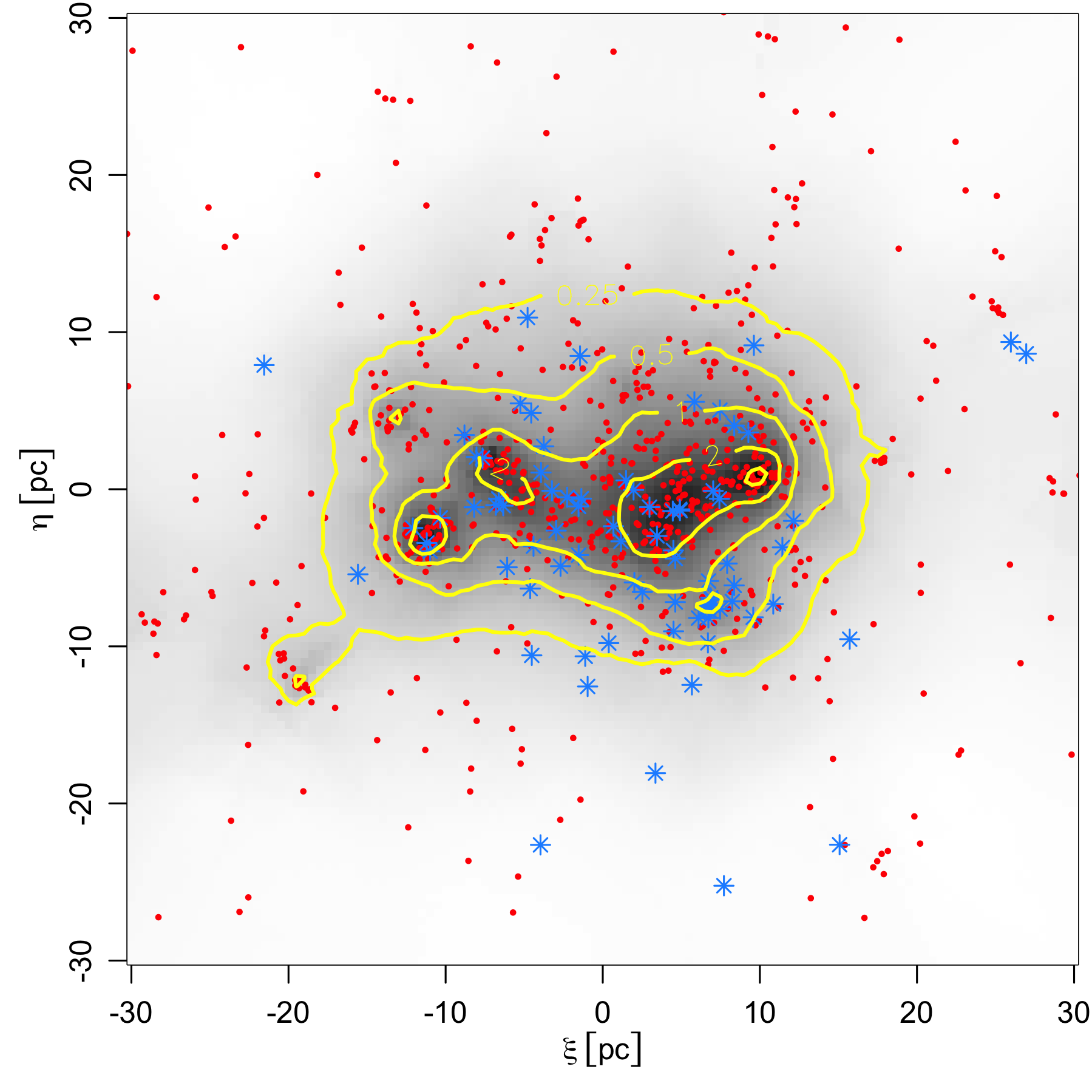}
        \includegraphics[width=0.48\textwidth]{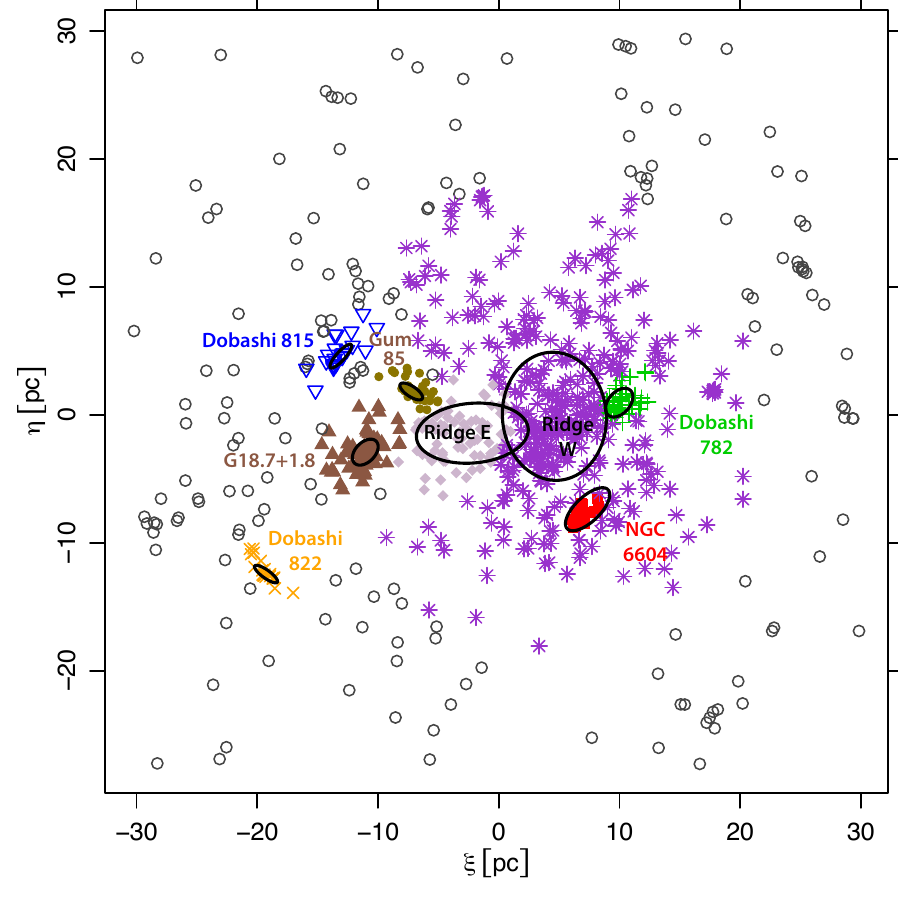}
    \caption{ 
    Left: Smoothed surface density distribution for members of Ser~OB2, in units of sample members per square parsec (greyscale image and yellow contour lines). The membership list is comprised of YSOs from the SPICY catalogue (red circles) and OB stars from the ALS III catalogue (blue asterisks), excluding those with inconsistent parallaxes.  Right: Mixture-model cluster solution with 8 cluster components. The black ellipses trace the isodensity contours of each component at 2 times the core radius. The assignments of stars to groups are indicated by the symbol shapes and colours. Distinct clusters are labelled. The two larger ellipses near the centre of the region identify the core (stellar members of the Ser~OB2 ridge W and E indicated by asterisks and diamonds, respectively). Ser~OB2 halo members are indicated by the open circles.}
    \label{fig:dens}
\end{figure*}

The OB stars and YSOs in our sample are concentrated around Galactic latitude $b\sim2^\circ$, roughly co-located with the mid-infrared nebulocity in the Spitzer/IRAC images. Although the OB stars and YSOs are mixed together, there appears to be some differences in their spatial distributions, which we compare with the Kolmogorov--Smirnov (K--S) test. This test shows a moderately statistically significant difference in the distribution of their Galactic longitudes ($p = 0.03$), which can be attributed to the OB stars being slightly more concentrated towards the centre of the OB association. However, the discrepancy is stronger in Galactic latitude ($p < 10^{-5}$), indicating that the OB stars preferentially have lower $b$. The massive cluster NGC~6604, containing several OB stars, is at the edge of the cloud, closer to the midplane than most of the rest of the association. Meanwhile, the YSOs are more strongly concentrated towards the middle of the cloud. 

To examine the spatial distribution of these sources, we convert positions in $(\ell, b)$ to a Cartesian $(\xi, \eta)$ coordinate system using the orthographic projection, where our $\xi$-axis is parallel to the direction of decreasing $\ell$ and our $\eta$-axis is parallel to the direction of increasing $b$ at the tangent point (Fig.~\ref{fig:dens}).\footnote{Our $\eta$ coordinate is approximately aligned with the Galactic $Z$-axis, increasing with height above the Galactic midplane.} In projection, the stars' spatial distribution is elongated in a direction parallel to the Galactic plane, with an aspect ratio of 2.7:1. The total length of this structure (encompasing 95\% of the probable members) is $\sim$53~pc. The clumpy spatial distribution of stars in the association is apparent from the adaptively smoothed surface-density map (Fig.~\ref{fig:dens}, left). This map was generated with the {\tt densityVoronoi} function in the R `spatstat' package \citep{BaddeleyRubakTurner2015,R-spatstat}, which estimates density from a Voronoi tessellation built on a randomly selected fraction of the points ($f=0.1$), uses the remaining points to estimate counts in each cell, and computes the mean surface density over 1000 repetitions \citep{OgataKatsuraTanemura2003,Ogata2004,Baddeley2007Validation}.

To identify clusters in the spatial distribution of Ser~OB2 association members, we used the mixture-model algorithm and R code from \citet{2014ApJ...787..107K}. The projected spatial distribution of cluster members is represented as a sum of multiple component distributions that capture the cluster structure. Young stellar clusters tend to be fairly extended and are best described by profiles with heavy wings. Thus, each component is modelled with a two-dimensional profile whose surface density is nearly uniform in a central core, has elliptical isodensity contours, and, at large distances from the centre, falls off as the inverse square of the elliptical radius. In sky coordinates $\mathbf{r} = (\xi,\eta)$, the surface density of a single component is
\begin{equation}
\Sigma(r)
= \Sigma_0 \left[1 + \frac{1}{r_c^2}(\mathbf{r} - \mathbf{r}_0)^\top A (\mathbf{r} - \mathbf{r}_0)\right]^{-1},
\end{equation}
where $\mathbf{r}_0$ is the centre, $\Sigma_0$ is the central surface density, $r_c$ is a core radius, and $A$ is a positive-definite $2\times 2$ matrix that sets the ellipticity and position angle. The final model is the sum of multiple such components, fit by maximum-likelihood estimation, with the number of components determined by minimisation of the Akaike Information Criterion \citep[AIC;][]{Akaike1974}, a likelihood-based statistic penalised by model complexity.

We find that a model with 8 clusters, roughly corresponding to the peaks in the smoothed surface-density map, provides the best fit ($AIC \approx 2366$) to the projected stellar distribution for Ser~OB2 (Fig.~\ref{fig:dens}, right). The residual map (not shown) contains both positive and negative residuals with typical amplitudes of $\pm 0.1$~pc$^{-2}$. The most significant residual of $+0.3$~pc$^{-2}$ lies between the east and west components of the central ridge, and may indicate that the model components do not fully describe this structure. Nevertheless, this residual is relatively small compared to the total stellar density $>$2~stars~pc$^{-2}$ in the same region. To evaluate the statistical evidence for each model component, we removed each component from the model in turn, refit the model, and recalculated the AIC. The $\Delta AIC$ changes range from 4 to 100, indicating that the additional complexity introduced by each component is justified by the improvement in fit.

The resulting model implies a centrally concentrated association, with a ridge-like overdensity in the core, represented by two components (Ser~OB2 ridge W and E), embedded in a more diffuse halo (Ser~OB2 halo). Superposed on this large-scale structure are several discrete clusters, each corresponding to a single component (Table~\ref{tab:subclusters}). These include NGC~6604 \citep{1786RSPT...76..457H}, a cluster associated with the bright Gum~85 nebulosity \citep{1955MmRAS..67..155G}, and clusters associated with the dark clouds Dobashi~782, 822, and 815 \citep{2011PASJ...63S...1D}, with Dobashi~782 also standing out as a bright patch of mid-infrared nebulosity in Spitzer images. An additional cluster, designated Ser~OB2 G18.7+1.8, lacks a previously catalogued counterpart, and is associated with a weak extinction feature visible in the Digitized Sky Survey~2 images \citep{1990AJ.....99.2019L}. Optical and infrared images reveal a compact subcluster around HD~167834 within the Ser~OB2 ridge W that our algorithm did not separate as a distinct component.

This substructure in Ser~OB2 is driven primarily by the YSO distribution rather than the OB stars. In our mixture-model components, OB stars are associated mainly with the central ridge, NGC~6604, and G18.7+1.8, while the other four clusters have cores devoid of OB stars. In contrast, the OB stars are distributed more smoothly across the association centre.

\begin{table*}
    \centering
    \caption{Mean Astrometric Motions of Ser~OB2 Subclusters}
    \label{tab:subclusters}
    \begin{tabular}{lrrrrrr}
        \hline
\multicolumn{1}{c}{Subregion} & \multicolumn{1}{c}{$\ell_0$}  & \multicolumn{1}{c}{$b_0$}  & \multicolumn{1}{c}{$n_\mathrm{samp}$} &  \multicolumn{1}{c}{$\bar \mu_{\ell^\star}$}   & \multicolumn{1}{c}{$\bar \mu_b$} & \multicolumn{1}{c}{$v_{LSR,K}$}   \\
\multicolumn{1}{c}{} & \multicolumn{1}{c}{(deg)}  & \multicolumn{1}{c}{(deg)}  & \multicolumn{1}{c}{} & \multicolumn{1}{c}{(mas~yr$^{-1}$)} & \multicolumn{1}{c}{(mas~yr$^{-1}$)} & \multicolumn{1}{c}{(km~s$^-1$)} \\
\multicolumn{1}{c}{(1)} & \multicolumn{1}{c}{(2)}  & \multicolumn{1}{c}{(3)}  & \multicolumn{1}{c}{(4)} & \multicolumn{1}{c}{(5)} & \multicolumn{1}{c}{(6)} & \multicolumn{1}{c}{(7)} \\
        \hline
Dobashi 782        & 18.15 & 1.94 &  11 & -2.33$\pm$0.08 & -0.60$\pm$0.10 & 29.7\\
NGC 6604           & 18.23 & 1.68 &   7 & -2.49$\pm$0.09 & -0.81$\pm$0.06 & \dots \\
Ser OB2 ridge W    & 18.33 & 1.87 & 144 & -2.15$\pm$0.03 & -0.56$\pm$0.03 & \dots  \\
Ser OB2 ridge E    & 18.55 & 1.85 &  45 & -2.03$\pm$0.06 & -0.71$\pm$0.05 & \dots \\
Gum 85            & 18.65 & 1.97 &  11 & -2.02$\pm$0.08 & -0.48$\pm$0.06 & 27.9\\
Ser OB2 G18.7+1.8 & 18.77 & 1.83 &  14 & -2.29$\pm$0.09 & -0.48$\pm$0.07 & 27.9\\
Dobashi 815        & 18.80 & 2.09 &   5 & -2.19$\pm$0.16 & -0.51$\pm$0.22 & 27.7\\
Dobashi 822        & 19.01 & 1.55 &   4 & -2.41$\pm$0.21 & -1.01$\pm$0.16 & 25.5\\
\hline
    \end{tabular}
	\flushleft{
	Column 1: Subregion designation. Column 2--3: Central location defined as the mid-range for $\ell$ and $b$ in each group. Column 4: Number of stars used in the astrometric analysis. Columns 5--6: Mean proper motions and standard errors for each group. Column 7: LSR velocity of peak $^{13}$CO $J=1$--0 emission in the associated molecular clump (Section~\ref{sec:cloud}). 
	}
\end{table*}

\begin{figure*}
    \includegraphics[width=0.48\textwidth]{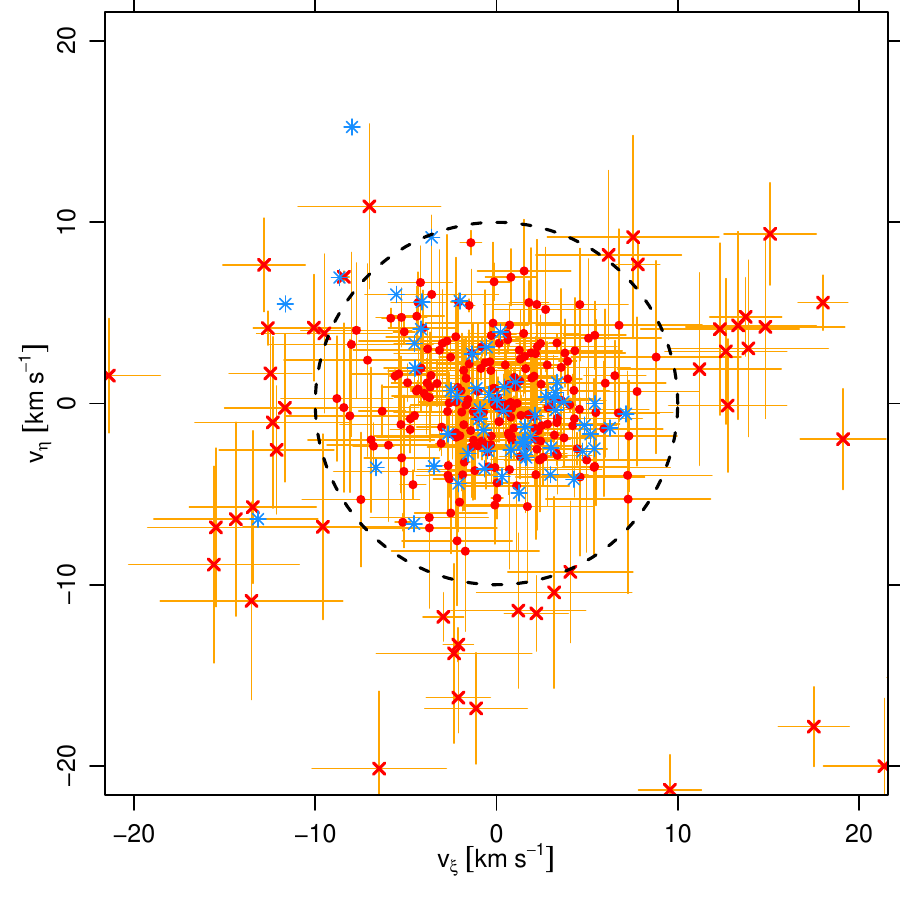}
    \includegraphics[width=0.48\textwidth]{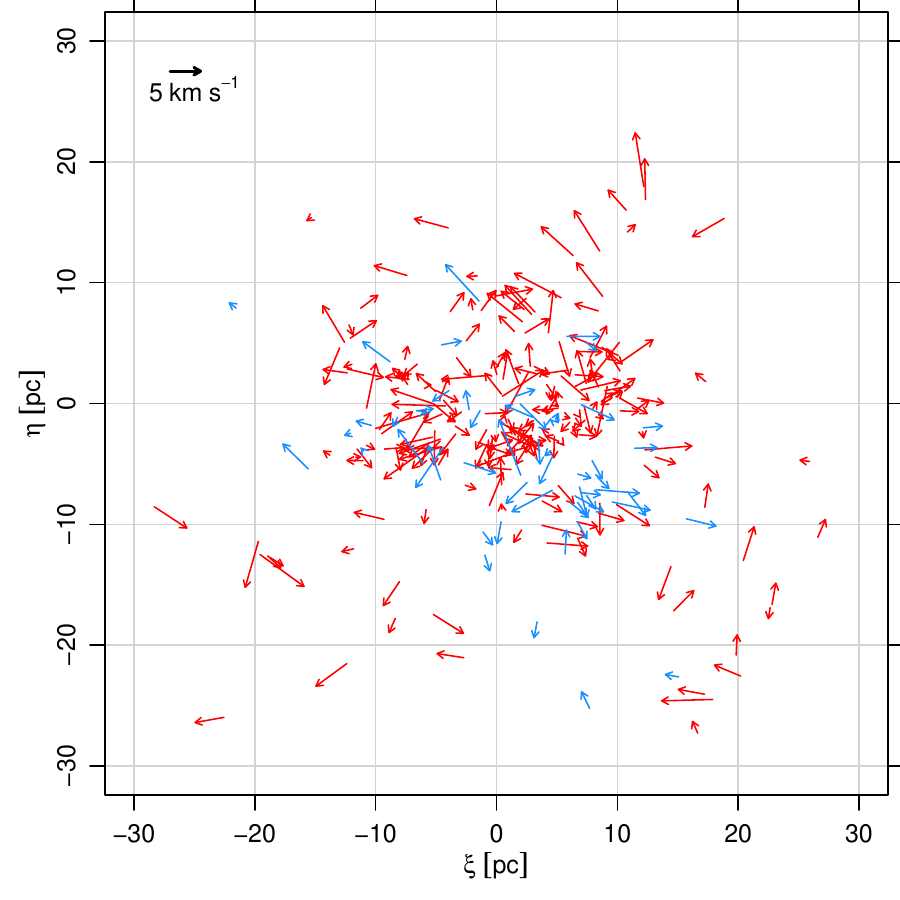}
    \caption{Left: The 2D projected velocity distribution for Ser~OB2, plotted with the same symbols as Fig.~\ref{fig:plx} and including only stars whose parallaxes are consistent with membership. We use $| \mathbf v | \leq 10$~km~s$^{-1}$ (dashed circle) as an approximate cut-off to separate the slow-moving cluster members from higher velocity stars, including walkaway and runaway association members and contaminants. 
    Right: Positions and velocities of OB stars (blue) and YSOs (red) from the $| \mathbf v | \leq 10$~km~s$^{-1}$ sample. The arrow lengths and directions indicate the projected velocities of these stars in the $(\xi, \eta)$ plane.
     }
    \label{fig:veldist}
\end{figure*}

\section{Kinematics}\label{sec:kin}

Proper motions $\mu_{\ell^\star}$ and $\mu_b$ are converted to plane-of-sky velocities $v_\xi$ and $v_\eta$ following \citet[][their Equations~1--4]{2019ApJ...870...32K}, including first-order corrections for perspective expansion and coordinate effects \citep{2009A&A...497..209V}. 
The systemic radial velocity of Ser~OB2 was estimated to be $v_{\mathrm{LSR},K}=27.3$~km~s$^{-1}$ \citep{SgrSpur} based on the $^{12}$CO $J=1$--0 map of \citet{2001ApJ...547..792D}, corresponding to a heliocentric velocity of $RV_{\rm helio}=12.5$~km~s$^{-1}$.\footnote{The kinematic Local Standard of Rest (LSRK) is defined by a fixed velocity of 20~km~s$^{-1}$ towards $\alpha=18^{\mathrm h}$, $\delta=+30^\circ$ (B1900.0) relative to the Solar System barycentre.} The maser G018.34+01.76 has a consistent radial velocity, $v_{\mathrm{LSR},K}=28\pm3$~km~s$^{-1}$ \citep{2019ApJ...885..131R}.

The velocity distributions of $v_\xi$ and $v_\eta$ 
have heavy tails and are better described by Student-$t$ distributions than by Gaussians. To estimate the location, scale $s$, and degrees of freedom $\nu$, we accounted for individual Gaia measurement uncertainties (assumed to be Gaussian) by convolving them with the intrinsic $t$ distribution and maximising the observed-data likelihood. For $v_\xi$ we obtained a velocity spread $s=3.5\pm0.6$~km~s$^{-1}$ with $\nu = 6\pm5$ for the OB stars and $s=3.1\pm0.2$~km~s$^{-1}$ with $\nu = 1.2\pm0.2$ for the YSOs. For $v_\eta$ we obtained $s=2.6\pm0.5$~km~s$^{-1}$ with $\nu = 3.2\pm1.7$ for the OB stars and $s=2.4\pm0.2$~km~s$^{-1}$ with $\nu=1.6\pm0.3$ for the YSOs. These results indicate similar spreads for the high- and low-mass populations, with slightly larger spreads parallel to the Galactic midplane than perpendicular to it. 

Figure~\ref{fig:veldist} (left) shows the 2D velocity distribution for Ser~OB2. A typical threshold for a walkaway star in a star-forming region is $\gtrsim$10~km~s$^{-1}$ \citep{2020MNRAS.495.3104S,2022MNRAS.510.1136P}. We find that 23\% of the YSOs and 7\% of the OB stars exceed this threshold. While these could be high-velocity members, potentially ejected by dynamical interactions, stars with discrepant velocities are also more likely to be field-star contaminants in the sample. Thus, these thresholds should be considered to be upper limits on the number of walkaways from Ser OB2 within the field of view. The remaining kinematic analysis focuses on the lower-velocity ($<$10~km~s$^{-1}$) members.

Figure~\ref{fig:veldist} (right) shows velocity vectors for the OB stars and YSOs, revealing complex velocity patterns. Although there are stars moving in all directions across the entire association, the pattern suggests common motions in subregions of the association.  

\begin{figure*}
    \includegraphics[width=1\textwidth]{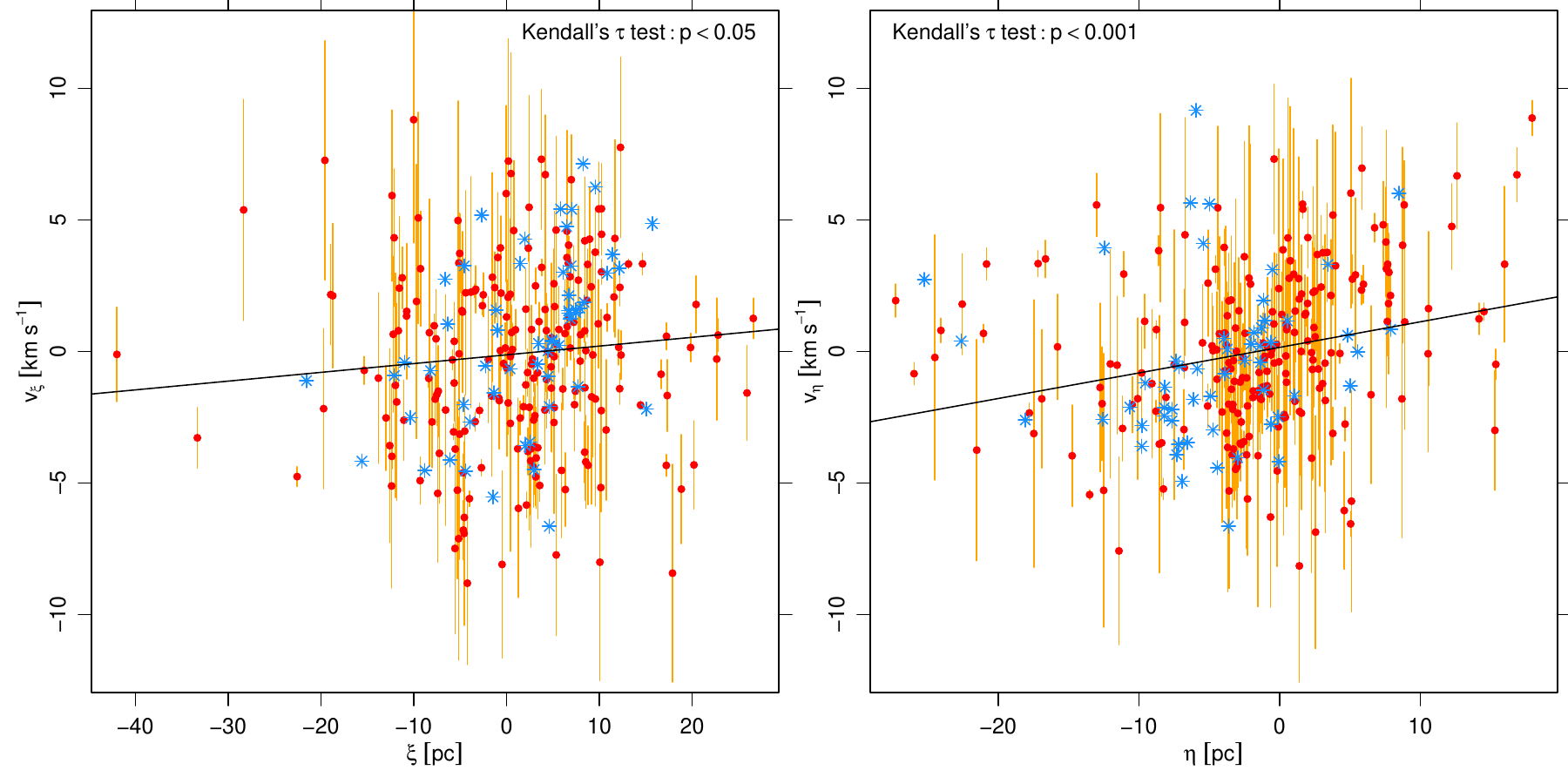}
    \caption{Position versus velocity for Ser~OB2 members along the $\xi$ and $\eta$ axes. The relations show significant scatter, attributable to both astrometric uncertainties and real velocity dispersions. Kendall's $\tau$ tests indicate positive correlations between positions and velocities, with $p = 0.005$ for $\xi$ (moderate statistical significance) and $p<10^{-6}$ (high statistical significance) for $\eta$. The OLS regression lines fit to the association's expansion gradient are shown. 
}
    \label{fig:expansion}
\end{figure*}

\subsection{Global expansion}\label{sec:expand}

Correlations between position and stellar velocity will be present if Ser~OB2 is undergoing global expansion or contraction (Fig.~\ref{fig:expansion}). Kendall’s $\tau$ rank correlation test indicates statistically significant positive correlations between $\xi$ and $v_\xi$ ($p = 0.005$) and between $\eta$ and $v_\eta$ ($p < 10^{-6}$). Linear fits using ordinary least-squares (OLS) regression yield mean velocity gradients of $0.03\pm0.02$~km~s$^{-1}$~pc$^{-1}$ along the $\xi$ axis\footnote{Given that the fitted slope of the $v_\xi$--$\xi$ relation is consistent with zero, it should be treated as a constraint on the expansion gradient rather than a detection. In this case, Kendall's $\tau$ test may be capturing velocity structure poorly fit by the linear model.} and $0.10\pm0.02$~km~s$^{-1}$~pc$^{-1}$ along the $\eta$ axis. 

Although the association is more spatially extended parallel to the Galactic plane than in the vertical direction, the velocity gradient is at least $\sim$3 times larger in $\eta$ (perpendicular to the midplane) than in $\xi$ (parallel to the plane). The detected $v_\eta$--$\eta$ velocity gradient accounts for only a small fraction of the total velocity dispersion (i.e., 6\% of the variance), but becomes apparent when the velocity field is examined across the full extent of the association.

\begin{figure*}
    \includegraphics[width=1\textwidth]{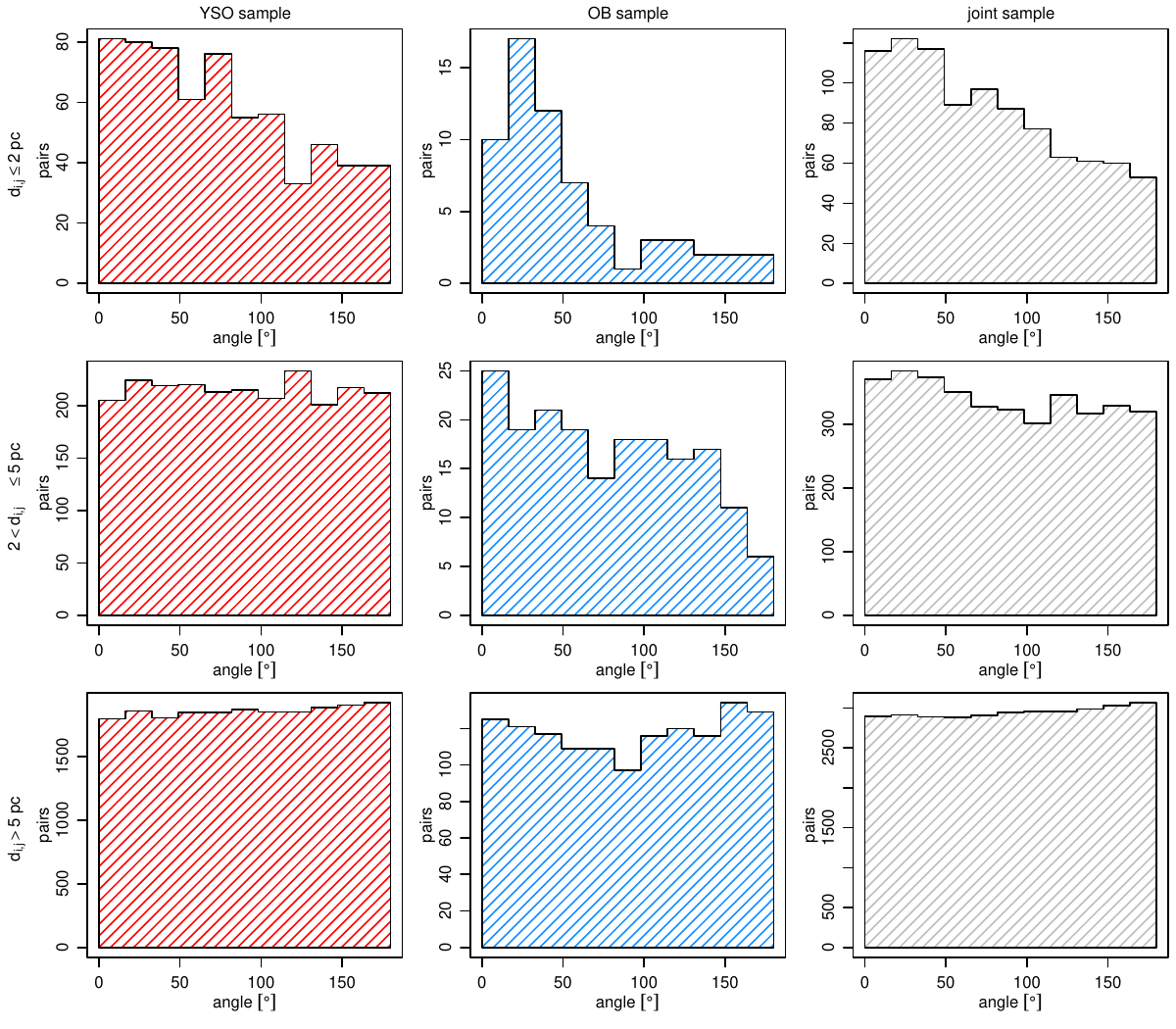}
    \caption{
Histograms of the pairwise angle $\theta_{i,j}$ between stellar velocity vectors, defined as $\theta_{i,j}=\cos^{-1}(\hat{v}_i\cdot\hat{v}_j)$, where $\hat{v}_k$ is the projected unit velocity of star $k$. Angles of $0^\circ$, $90^\circ$, and $180^\circ$ correspond to parallel, orthogonal, and antiparallel motion, respectively. Columns (left to right) show the YSO, OB, and joint samples. Rows (top to bottom) correspond to projected pair separations $d_{i,j}\le 2$~pc, $2<d_{i,j}\le 5$~pc, and $d_{i,j}>5$~pc. A pronounced excess near $0^\circ$ is evident in the $d_{i,j}\le 2$~pc bin for all samples, indicating aligned motions at small separations. At larger separations the distributions show no preferred alignment.    
     }
    \label{fig:aligned}
\end{figure*}

\begin{figure*}
    \includegraphics[width=1\textwidth]{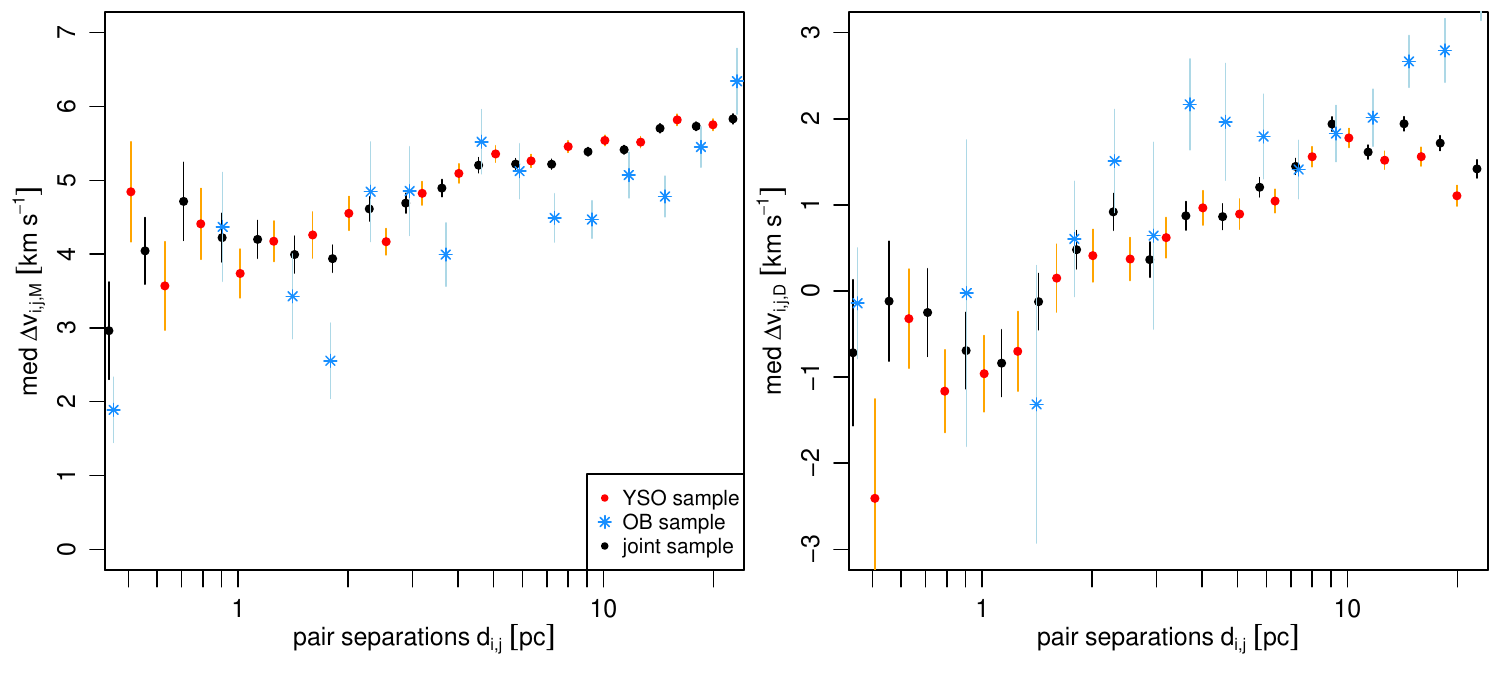}
    \caption{
Median values for the VSAT statistics, $\Delta v_{ij,M}$ (left panel) and $\Delta v_{ij,D}$ (right panel), binned in projected pair separations $d_{i,j}$. Error bars indicate 1$\sigma$ standard error estimates. Both statistics show an overall rise in median value with greater pair separation. Frequentist tests, comparing observed results to simulations where velocities were randomly permuted among the stars, show that trends for all samples are strongly statistically significant ($p<0.001$).}
    \label{fig:VSAT}
\end{figure*}

\subsection{Correlated stellar velocities}\label{sec:vsat}

To characterise the kinematic structure of Ser~OB2 over a range of spatial scales, we use pairwise velocity statistics in a structure-function-style analysis.

To test the alignment of velocities, we examined pairs of stars and calculated the angle between their velocity vectors. This angle is defined
\begin{equation}
\theta_{ij} = \arccos \frac{\mathbf{v}_i \cdot \mathbf{v}_j}{|\mathbf{v}_i||\mathbf{v}_j|},
\end{equation}
where $\mathbf{v}_i$ and $\mathbf{v}_j$ are the velocity vectors of the two stars as seen in projection in the rest frame of the association. Fig.~\ref{fig:aligned} shows histograms of these angles for pairs of points over a range of separations. We consider pairs of YSOs, pairs of OB stars, and pairs of the joint sample containing both YSOs and OB stars. When pairs are separated by a projected distance less than or equal to 2~pc, there is a clear preference for nearly parallel velocity vectors ($\theta_{ij}$ near 0$^\circ$) and a deficit of nearly anti-parallel velocity vectors ($\theta_{ij}$ near 180$^\circ$). This is apparent for the YSO, the OB stars, and the joint sample. The preference is strongest for the OB stars, which may be explained by smaller velocity uncertainties for this subset. The preference for parallel velocity vectors is weaker when considering stars with projected separations between 2 and 5~pc, and is only clearly present for the OB star sample. When considering pairs with projected separations greater than 5~pc, there is no significant correlation in velocities. This effect cannot be an artefact of the correlated uncertainties in Gaia proper motions because the velocity dispersion of the sample ($\sim$3~km~s$^{-1}$) is much larger than the maximum amplitude of the correlated uncertainties ($\pm$0.2~km~s$^{-1}$).

\citet{2019MNRAS.483.3894A} proposed two velocity-structure-analysis-tool (VSAT) statistics for OB associations,
\begin{equation}
\Delta v_{ij,M} =  |\mathbf{v}_i - \mathbf{v}_j  |
\end{equation} 
and 
\begin{equation}
\Delta v_{ij,D} =  \frac{(\mathbf{r}_i - \mathbf{r}_j)\cdot(\mathbf{v}_i - \mathbf{v}_j)}{|\mathbf{r}_i - \mathbf{r}_j|},
\end{equation}
where $\mathbf{v}_i$ is the projected 2D velocity of the $i$-th star as above, and $\mathbf{r}_i$ is its 2D projected position vector. The first of these statistics provides the difference in velocities, while the second is the component of the velocity difference parallel to the separation between the stars. 

Figure~\ref{fig:VSAT} shows the median values of $\Delta v_{ij,M}$ and $\Delta v_{ij,D}$ in bins of pair separation for the OB, YSO, and combined samples. In all samples, for separations $\gtrsim 2$~pc, both statistics tend to increase with separation. The OB sample shows larger departures from a strictly monotonic increase, likely due to its smaller size and correspondingly larger stochastic fluctuations. At separations $\lesssim$2~pc the trend appears to break down. In particular, Gaia measurement uncertainties may impose a floor on the median $\Delta v_{ij,M}$, plausibly explaining why this statistic does not continue to decrease towards the smallest separations.

The increasing median values of $\Delta v_{ij,M}$ and $\Delta v_{ij,D}$ between 2 and 20~pc show that more widely separated stars tend to have larger velocity differences. Since $\Delta v_{ij,D}$ is defined such that positive values correspond to velocities that increase the separation between stars, its behaviour further indicates that these motions are preferentially divergent. This pattern is consistent with an expanding association in which the expansion velocity grows with distance. At larger separations, the median $\Delta v_{ij,D}$ tends to be higher for OB stars than for the YSOs, suggesting a slightly more pronounced expansion signature among the OB stars.

The standard error of each binned median (Fig.~\ref{fig:VSAT}) was estimated as $0.93\,\mathrm{IQR}/\sqrt{n}$, derived for approximately normally distributed data, where IQR is the interquartile range and $n$ is the number of data points per bin. Because the VSAT statistics are based on pairs of points, values are not strictly independent, so these standard errors should be interpreted with caution. To rigorously assess the statistical significance of the trends, we used simulation-based hypothesis testing. For this test, the null hypothesis is that stellar position and velocity are independent, which we simulate by randomly permuting the stellar velocity vectors among the stars. For the test statistic, we used the slope of the linear OLS fit to the median~VSAT--$\log d_{i,j}$ relation over the range $0.4\leq d_{i,j}\leq23$~pc (as shown in Fig.~\ref{fig:VSAT}). For each test, we ran 10,000 simulations, and computed the one-sided $p$-value as the fraction of simulated slopes at least as extreme as the observed slope in the observed direction. For the YSO sample and the joint sample, we obtained $p<0.001$ (high significance) for the trends in both VSAT statistics. For the OB sample, we obtained $p\approx0.002$ and $p\approx 0.01$ (moderate significance) for the trends in $\Delta v_{ij,M}$ and $\Delta v_{ij,D}$, respectively. 

\subsection{Subcluster Kinematics}\label{sec:velsubcl}

\begin{figure}
    \includegraphics[width=0.5\textwidth]{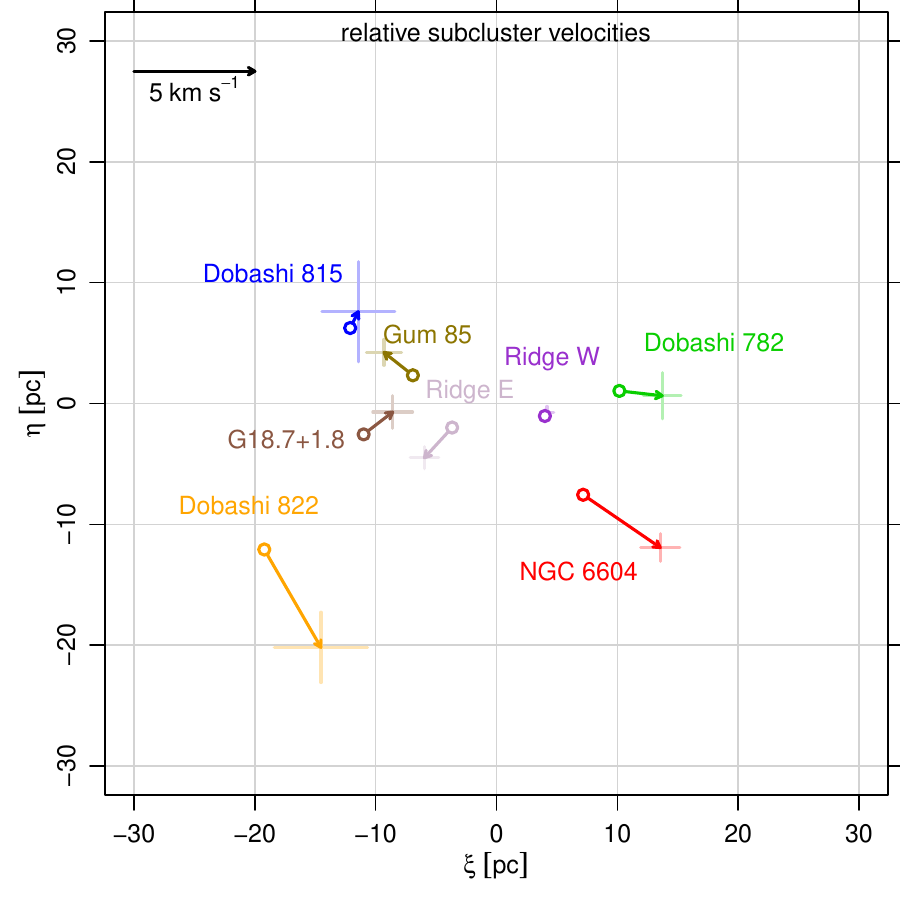}
    \caption{Positions and velocities of the subclusters within Ser~OB2. Arrows indicate the mean velocity vector of each subcluster, using the scale shown in the upper right corner of the plot. The error bars indicate standard errors on the mean velocities.}
    \label{fig:subcl_vel}
\end{figure}

To investigate the relative motions of the clusters within Ser~OB2 (Table~\ref{tab:subclusters}), we computed the mean velocity of the stars in each group. The resulting subcluster velocities, measured relative to the mean systemic velocity, range from $\sim$0~km~s$^{-1}$ to $\sim$5~km~s$^{-1}$, and the differences between groups are statistically significant. However, they do not form a simple pattern of global coalescence or divergence (Fig.~\ref{fig:subcl_vel}). In the central region, the two groups comprising the ridge (Ser~OB2 ridge E and W) are moving apart, and several surrounding subclusters (NGC~6604, Dobashi~782, and Gum~85) are also moving away from the centre of the association. In contrast, G18.7+1.8 is moving inward, while Dobashi~815 and 822 move in orthogonal directions. Taken together, these motions point to a complex internal velocity field rather than a single coherent pattern.

\section{Molecular Cloud}\label{sec:cloud}

\begin{figure*}
    \includegraphics[width=0.75\textwidth]{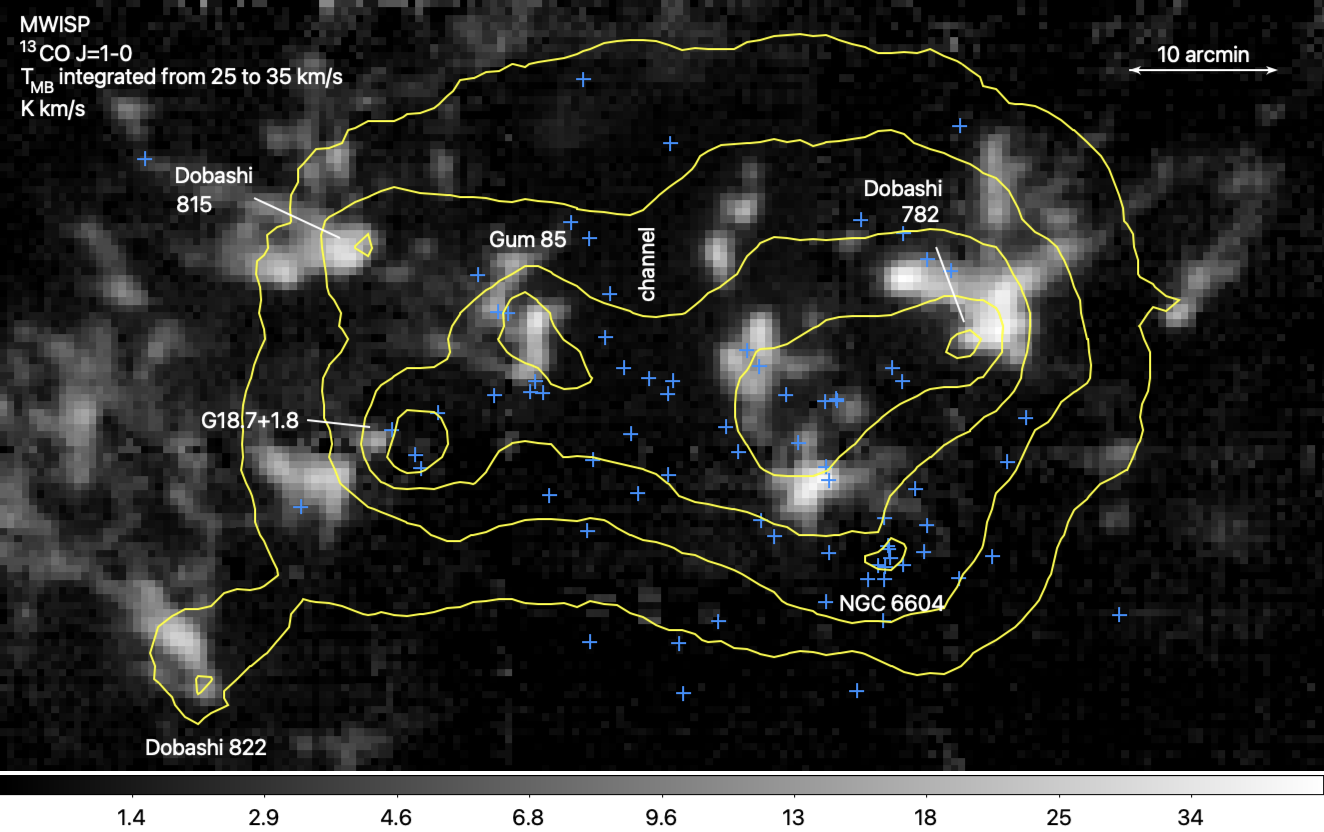}
    \caption{The MWISP $^{13}$CO $J=1$--0 zeroth moment map between 25 and 35~km~s$^{-1}$. The greyscale image shows integrated emission in units of $\mathrm{K}\cdot\mathrm{km}$~s$^{-1}$. OB stars are marked by blue crosses, Ser~OB2 clusters (Table~\ref{tab:subclusters}) are labelled, and the stellar surface density contours from Fig.~\ref{fig:dens} are drawn in yellow. The molecular-gas-free channel through the centre of the association, also visible in the Spitzer image, is labelled.}
    \label{fig:mwisp}
\end{figure*}

The Milky Way Imaging Scroll Painting Survey (MWISP) Data Release~1 \citep{2019ApJS..240....9S,2025arXiv251208260Y} provides molecular-gas maps from the Purple Mountain Observatory 13.7-m telescope covering the Ser~OB2 region. In the $^{13}$CO $J=1$--0 map (Fig.~\ref{fig:mwisp}), dense molecular clumps are distributed across the whole association. 

Distinct cloud clumps are associated with Ser~OB2 subclusters in five regions: Gum~85, Dobashi~782, 815, 822, and G18.7+1.8. For each clump, we record the $v_{LSR,K}$ corresponding to the velocity channel of peak $^{13}$CO emission (Table~\ref{tab:subclusters}). In several cases (e.g.\ Dobashi~782, 815, and G18.7+1.8), the clumps show sharper centre-facing edges in the $^{13}$CO image and more diffuse emission extending away from the centre, with the subclusters on the centre-facing side. These morphologies resemble molecular clumps exposed to external irradiation at the edges of expanding H\,{\sc ii} regions, and are commonly attributed to externally driven surface erosion \citep[e.g.,][]{2013A&A...556A.105O}. The cluster NGC~6604, rich with OB stars, has a larger projected separation from the nearest dense core. 

In the centre of the association, a molecular-gas-free channel extends vertically (in Galactic coordinates) through the cloud. The channel is also visible as a vertical gap in the nebulosity in the Spitzer/IRAC images (Fig.~\ref{fig:irac}), with walls traced by polycyclic aromatic hydrocarbon (PAH) emission. Several OB stars lie in projection within this channel, suggesting they helped ionise the H\,{\sc ii} region and clear a passage through the cloud. Because many of the OB stars (including those in NGC~6604) lie below the cloud in Galactic latitude while the thermal chimney extends upward, this channel may have served as the pathway where the H\,{\sc ii} region broke through the cloud and inflated the chimney.

\section{Discussion}\label{sec:discussion}

As one of the youngest and richest OB associations within 2~kpc of the Sun, Ser~OB2 has provided a unique opportunity to examine the kinematics of an OB association at a very early stage. In contrast, other systems like Sco--Cen, Sco~OB1 and Vela~OB2 have stellar populations ranging from $\gtrsim$5--20~Myr in age and may only contain isolated pockets of ongoing star formation \citep[e.g.][]{2016MNRAS.461..794P,2009MNRAS.393..538J,2018A&A...615A.148D,2019A&A...626A..17C,2020NewAR..9001549W,2023A&A...678A..71R}.
Like these other OB associations, the relatively large internal velocity dispersions (2.4--3.5~km~s$^{-1}$) and diameter ($\sim$50~pc) of Ser~OB2 means that it is unlikely to contain sufficient mass to be globally gravitationally bound, which would require a cloud mass exceeding $10^5$~$M_\odot$ \citep[][their Equation~4]{2010ARA&A..48..431P}. Individual clusters within Ser~OB2 may be gravitationally bound, but our sample is insufficient to test this.

We observe distinct velocity patterns on different spatial scales, and turbulence provides a natural framework for interpreting this structure. In a turbulent molecular cloud, velocity differences increase with spatial scale \citep{2004RvMP...76..125M}, which could be reflected in the velocity statistics of the stars they form \citep{2020MNRAS.495.3474A}. On the smallest scales probed by our sample ($\lesssim2$~pc), the $\theta_{ij}$ statistic indicates that nearby stars have preferentially aligned velocity directions. These scales encompass typical sizes of individual young clusters and subclusters \citep{2003ARA&A..41...57L,2014ApJ...787..107K,2017AJ....154..214K}, whose stars may have formed in the same molecular clump with similar initial velocities. The VSAT statistics probe pairwise velocity differences over more than an order of magnitude in scale. For separations of $\sim$2--20~pc, both statistics show increasing velocity differences with separation, reminiscent of the \citet{1981MNRAS.194..809L} relations where cloud velocity dispersions increase with length scale. The VSAT $\Delta v_{ij,D}$ statistic has a positive median for separations $>$2~pc, indicating net divergent motions, with divergence increasing with spatial separation. On intermediate scales, several clusters are moving outward, but others deviate from this trend, implying a more complex velocity field. On the largest scales, we detect gradual global expansion of the association along the axis perpendicular to the Galactic plane. 

It is plausible that the asymmetric velocity gradient could result from stochastic turbulent fluctuations in a globally unbound system. Nevertheless, it is notable that the gradient is oriented perpendicular to the Galactic plane, where the relative placement of the OB stars, molecular gas, and the thermal chimney provides a natural direction. This feedback geometry may induce an asymmetric expansion of the cloud which could be imprinted on the stars via sequential star formation.

Previous work has shown that the $\sim$4--5~Myr-old OB population in Ser~OB2 is energetically capable of driving the large H\,{\sc ii} region Sh~2-54 and the associated `Stockert chimney' \citep[][]{1987Ap&SS.136..281K,2000AJ....120.2594F}.\footnote{An expanding bubble model has been used to successfully describe another thermal chimney in the W4 region, obtaining an expansion age of 2.5~Myr \citep{1999ApJ...516..843B}. While both chimneys are of similar length, the Ser~OB2 chimney is much narrower and its ionising source is less certain \citep{1999ASPC..168..287N}.} Although the interface between the Sh~2-54 H\,{\sc ii} region and the molecular cloud is likely to be complex (Section~\ref{sec:cloud}), the OB stars are preferentially nearer the Galactic plane, so the net force of the expanding H\,{\sc ii} region is likely to be directed upward (i.e., in the $+\eta$ direction). If the expansion of the Sh~2-54 H\,{\sc ii} region has been pushing the cloud away from the midplane, stars forming in this accelerated gas would inherit upward velocities, producing a vertical velocity gradient.

If the H\,{\sc ii} region is overpressurised with $p\sim10^{-10.5}$~dyn~cm$^{-2}$ \citep{1987Ap&SS.136..281K}, and we assume a molecular-cloud mass $M_\mathrm{cl}\sim5\times10^4$~$M_\odot$ (for a star-formation efficiency $\epsilon=0.1$) and a $\sim$50~pc-diameter interface between H\,{\sc ii} region and cloud, the implied acceleration is
$a \approx \pi R^2 p/M_\mathrm{cl} \approx 1.9$~km~s$^{-1}$~Myr$^{-1}$. To gauge whether an accelerating cloud can imprint the observed spatial velocity gradient, we built a simple toy model in which stars form at a constant rate in such an accelerating cloud. In the model, stars inherit the cloud's instantaneous mean position and velocity plus Gaussian scatter of 5~pc and 1~km~s$^{-1}$, respectively. After formation, stars move ballistically at fixed velocity. Fitting velocity vs.\ position with OLS yields a gradient of $0.10\pm0.01$~km~s$^{-1}$~pc$^{-1}$ after 2.5~Myr, which is similar to the observed gradient, although uncertainties in the acceleration duration and initial dispersions limit more quantitative modelling.

Several clusters from the \citet{2023A&A...673A.114H} catalogue lie in projection near Ser~OB2, including CWNU~367, NGC~6604, CWNU~493, Theia~1656, UBC~344, and UBC~1019, but only NGC~6604 and UBC~344 (the Ser~OB2 ridge) overlap in membership with our Ser~OB2 sample. CWNU~367 ($d=1368$~pc) and CWNU~493 ($d=1613$~pc) are foreground and have older estimated ages. Theia~1656 ($d=2070$~pc) lies $\sim$0.5$^\circ$ closer to the Galactic midplane and has an age of 13~Myr, while UBC~1019 ($d=1897$~pc) is at a similar distance and just above Ser~OB2 in Galactic coordinates but is much older ($\sim$200~Myr). The \citet{2023A&A...673A.114H} catalogue's recovery of Ser~OB2 members was likely limited by the challenge of identifying distant YSO clusters using Gaia alone, given that YSOs tend to have lower-quality Gaia astrometry and higher velocity dispersions than disc-free stars in older clusters. 

\section{Conclusion}\label{sec:conclusion}

Ser~OB2 is a large OB association notable for its youth ($\lesssim 5$~Myr), its moderate height above the Galactic midplane (65~pc), and its role in launching a $\sim$200~pc thermal chimney. We have analysed a sample of Ser~OB2 members assembled from OB stars in the ALS~III catalogue and YSOs from the SPICY catalogue, with membership vetted using Gaia~DR3 parallaxes. Owing to the inclusion of YSOs, this is the first study of the region to incorporate a substantial sample of low- and intermediate-mass members. Although spatial distributions differ moderately, the OB stars and YSOs show similar kinematics, with increasingly divergent motions at larger separations.

The main results of this study are as follows:
\begin{itemize}
\item We identified 60 OB-star and 284 YSO probable association members, whose Gaia parallaxes are consistent with a single association distance of $1950\pm30$~pc. In addition, we retained 21 OB-star and 534 YSO member candidates lacking reliable Gaia astrometry. Scaling the OB population to a standard IMF implies a total association membership of $\sim$10$^4$ stars (Section~\ref{sec:mem}).
\item While previous work indicated an age of 4--5~Myr for the OB population \citep[e.g.][]{2000AJ....120.2594F}, the same age would imply implausibly large extinctions for the low-mass YSOs. The YSOs are more consistent with an age of $\sim$1~Myr, implying an overall age spread of a few Myr in the association (Section~\ref{sec:isochrone}).
\item The Ser~OB2 members exhibit a clumpy spatial distribution elongated parallel to the Galactic midplane. Although the spatial distributions of OB stars and YSOs overlap, the OB stars are preferentially closer to the Galactic midplane and more concentrated toward the centre of the association in Galactic longitude (Section~\ref{sec:clust}). In this central region, which contains many of the O stars, we identify a gap in the molecular clouds that may serve as a channel for launching the thermal chimney (Section~\ref{sec:cloud}).
\item We identified six discrete stellar clusters (NGC~6604, Gum~85, Dobashi~782, 815, 822, and G18.7+1.8) superimposed on a relatively smooth, centrally concentrated stellar distribution (Section~\ref{sec:clust}). Five of these clusters are associated with distinct molecular cloud clumps in $^{13}$CO $J=1$--0 MWISP data cubes. 
\item After converting Gaia proper motions to plane-of-sky velocities, we find characteristic one-dimensional velocity spreads of 2.4--3.5~km~s$^{-1}$ (Section~\ref{sec:kin}). For the remaining kinematic analysis we excluded sources with $|\mathbf{v}|>10$~km~s$^{-1}$, as velocity outliers are more likely to be ejected walkaway/runaway stars or contaminants and are unlikely to trace the association's bulk motions.
\item Pairwise velocity statistics show that stellar velocities are correlated on spatial scales $\lesssim$2~pc, but velocity differences increase for larger separations. Velocities become increasingly divergent with increasing spatial separations (Section~\ref{sec:vsat}). Projected subcluster velocities differ from the association mean by up to 5~km~s$^{-1}$ and display a complex pattern of motions (Section~\ref{sec:velsubcl}).
\item The association exhibits a gradual but highly statistically significant global expansion perpendicular to the Galactic midplane, with a gradient of $0.10\pm0.02$~km~s$^{-1}$~pc$^{-1}$ (Section~\ref{sec:expand}). This vertical expansion gradient may reflect sequential star formation, if the cloud has been accelerated by the expanding Sh~2-54 H\,{\sc ii} region that also drives the Stockert chimney (Section~\ref{sec:discussion}).
\end{itemize}

Large OB associations are known to be expanding, with stellar densities too low for them to remain bound \citep{1964ARA&A...2..213B,2018PASP..130g2001G}. However, the physical processes that set their kinematics remain debated. Gaia shows that most young clusters---including those nested within larger associations---expand once they become optically visible \citep{2019ApJ...870...32K}, consistent with gas expulsion \citep{1983ApJ...267L..97M,2000ApJ...542..964A}, although alternative explanations exist (Section~\ref{sec:discussion}). Whether the same mechanisms dominate on association scales is less clear. Detailed Gaia studies of older (tens of Myr) systems, including Sco--Cen \citep{2025arXiv250913607H} and Vela~OB2 \citep{2019A&A...626A..17C}, link present-day expansion to turbulence and feedback imprinted in natal clouds. Because many nearby associations are in later evolutionary stages where progenitor clouds have largely dispersed \citep{2007prpl.conf..345B}, it has been challenging to directly connect their current kinematics to the star-formation conditions.

Our analysis of Ser~OB2 reveals strong expansion signatures while star formation is ongoing (Section~\ref{sec:kin}). Expansion dominates the velocity field on scales $>$2~pc up to the full $\sim$50~pc extent of the system, but the flow departs from simple homologous expansion. The expansion gradient is larger perpendicular to the Galactic midplane than parallel to it. Ser~OB2 is also spatially clumpy, with correlated motions within clusters and partially random relative motions between subclusters, likely reflecting the kinematics of molecular clumps in a turbulent cloud (Section~\ref{sec:cloud}).

The disruption of clusters within Ser~OB2 may contribute to the region's overall expansion. However, cluster dispersal alone is unlikely to explain expansion on association scales, because it operates on a crossing timescale \citep{2006MNRAS.373..752G}. For a characteristic vertical size of $\sim$30~pc and a velocity dispersion of $\sim$2.4~km~s$^{-1}$ in this direction, the crossing time is $\gtrsim$10~Myr, greater than the ages of most members.
 
Alternatively, the large-scale expansion pattern may be an imprint of the velocity field in the molecular cloud from which the stars formed. Large-scale velocity patterns in the gas may arise from turbulent fluctuations \citep{2004RvMP...76..125M} or from the expansion of a molecular-cloud shell driven by feedback \citep{1977ApJ...214..725E}. In Ser~OB2, the configuration of the H\,{\sc ii} region and molecular cloud would favour expansion away from the Galactic plane, in the same orientation as the association-scale stellar velocity gradient. Furthermore, the difference in the mean ages of the OB stars and YSOs is consistent with the sequential star-formation scenario of \citet{1977ApJ...214..725E}. These results suggest that the expansion and divergence patterns seen in young stellar populations need not arise solely from gas expulsion, but may reflect the initial kinematics of stars upon formation.

% My God, it's full of stars!

\section*{Acknowledgements}

We thank the referee for insightful perspectives that helped strengthen this work. This research used the NASA/IPAC Infrared Science Archive (IRSA), funded by NASA and operated by Caltech; archival data from the Spitzer Space Telescope, operated by JPL under a NASA contract; and data from ESA's Gaia mission, processed by the Gaia Data Processing and Analysis Consortium (DPAC), with funding from national institutions, particularly those participating in the Gaia Multilateral Agreement. This work made extensive use of SAOImage DS9 \citep{2003ASPC..295..489J}, the CDS services \citep{2000A&AS..143....9W,2000A&AS..143...23O,2000A&AS..143...33B}, topcat/stilts \citep{2005ASPC..347...29T,2025arXiv250103299T}, and the R language \citep{R2022}. This work used NASA’s Astrophysics Data System (ADS) under Cooperative Agreement 80NSSC25M7105 and NASA’s Science Explorer (SciX) under Cooperative Agreement 80NSSC21M00561\citep{2000A&AS..143...41K,2023AGUFMSH33C3072G}. 

For the purpose of open access, the authors have applied a Creative Commons Attribution (CC BY) license to any Author Accepted Manuscript version arising from this submission.

\section*{Data Availability}

The SPICY catalogue is available via NASA/IPAC IRSA \citep[DOI: 10.26131/IRSA556;][]{2021ipac.data.I556K}.
Gaia DR3 data are available from the ESA Gaia Archive \citep[DOI: 10.5270/esa-1ugzkg7;][]{GaiaCollaboration2016}. The ALS~III catalogue is available from the ALS portal \citep[\url{https://als.cab.inta-csic.es};][]{2025MNRAS.543...63P}. 
 
\bibliographystyle{mnras}
\bibliography{ms.bbl} 

\bsp	\label{lastpage}
\end{document}